\documentstyle[prd,aps]{revtex}

\begin{document}
\twocolumn[\hsize\textwidth\columnwidth\hsize\csname@twocolumnfalse\endcsname
\preprint{NORDITA-2001-???-HE}
\title{The Glueball sector of two-flavor Color
Superconductivity}
\author{Rachid {\sc Ouyed}
\quad and \quad Francesco {\sc Sannino}
}
\address{NORDITA, Blegdamsvej 17
DK-2100 Copenhagen \O, Denmark}

\date{March 2001}
\maketitle

\begin{abstract}

We construct the effective Lagrangian describing the light
glueballs associated with the unbroken and confining $SU_c(2)$
color subgroup for the 2 flavor superconductive phase of QCD. This
Lagrangian constitutes a key ingredient for understanding the non
perturbative physics of 2 flavor color superconductivity. We
estimate the two photon decay process of the light glueballs using
the saturation of the electromagnetic trace anomaly at the
effective Lagrangian level.
 The present results
are particularly relevant to our model of Gamma Ray Bursts based
on color superconductivity in Quark Stars (R.~Ouyed and F.~Sannino
astro-ph/0103022).

\end{abstract}
\vskip 2cm
 ]

\section{Introduction}
\label{uno}

Quark matter at very high density  is expected to behave as a
color superconductor \cite{B,BL}. Recent work had lead to a
renewed interest on the subject \cite{ARW,RSSV,RW}. This phase is
characterized by its gap energy ($\Delta$) associated to
quark-quark pairing. In such a phase, the color symmetry is
spontaneously broken and a hierarchy of scales, for given chemical
potential, is generated. Indicating with $g_s$, the underlying
coupling constant, the relevant scales are: the chemical potential
$\mu $ itself, the dynamically generated gluon mass $m_{gluon}\sim
g_s\mu$ and $\Delta$. Since for high $\mu$ the coupling constant
$g_s$ (evaluated at a fixed scale $\mu $) is $\ll 1$, we have:
\begin{equation}
\Delta \ll m_{gluon}\ll \mu \ .
\end{equation}
The low-energy effective Lagrangian describing the in medium
fermions and the broken sector of the $SU_c(3)$ color groups for
the 2 flavor color superconductor (2SC) has been constructed in
Ref.~\cite{CDS}. The 3 flavor case (CFL) has been developed in
\cite{CG}. The effective theories describing the electroweak
interactions for the low-energy excitations in the 2SC and CFL
case can be found in \cite{CDS2001}. In Reference \cite{rischke2k}
it has been shown that the confining scale of the unbroken
$SU_c(2)$ color subgroup is lighter than the superconductive gap
$\Delta$. This is a consequence of the high dielectric constant
$\epsilon$ of the 2SC medium \cite{rischke2k}. The unbroken
$SU_c(2)$ theory is still confining and the light glueball like
particles are expected to be light with respect to $\Delta$ and
hence play a relevant role at low energies.


In this paper we generalize the effective Lagrangian for 2 flavor
by properly taking into account the confined degrees of freedom
associated with the low energy unbroken $SU_c(2)$ gauge
interactions. We estimate the two photon decay process of the
light glueballs using the saturation of the electromagnetic trace
anomaly at the effective Lagrangian level. The glueball decay was
found to be crucial in our model for powering Gamma Ray Bursts
\cite{OS} (involving a 2SC layer at the surface of Quark Stars).
It led to establish a link between QCD at high matter density and
Gamma Ray Bursts observables (energy and duration) extracting
vital information about the QCD phase diagram. In particular a
value for the critical temperature (of the order of $15$~MeV)
above which color superconductivity cannot exist has been
determined for densities few times nuclear matter density
\cite{OS}.

In Section \ref{review} we briefly review, while setting our
conventions, the low-energy effective Lagrangian for the 2SC phase
of QCD \cite{CDS,CDS2001}. The latter describes the, in medium,
fermions and the broken $SU_c(3)$ gluon sector. In Section
\ref{Glueball} we construct the effective Lagrangian describing
the light glueballs associated with the unbroken $SU_c(2)$ color
subgroup by using the information inherent to the trace anomaly
and the medium effects related to a non-vanishing dielectric
constant \cite{rischke2k}. In \ref{decay} we estimate the, in
medium, glueball to two photon decay process which is relevant to
our model of Gamma Ray Bursts Ref.~\cite{OS}. We conclude in
\ref{conclusion}.

\section{2SC Effective Lagrangian review}
\label{review}

 QCD with 2 flavor has gauge symmetry $SU_{c}(3)$
and global symmetry
\begin{equation}
SU_{L}(2)\times SU_{R}(2)\times U_{V}(1)\ .
\end{equation}
\noindent  At high matter density a color superconductive phase
sets in and the associated diquark condensates leaves invariant
the following symmetry group:
\begin{equation}
\left[ SU_{c}(2)\right] \times SU_{L}(2)\times SU_{R}(2)\times
\widetilde{U}_{V}(1)\ ,
\end{equation}
where $\left[ SU_{c}(2)\right] $ is the unbroken part of the gauge
group. The $\widetilde{U}_{V}(1)$ generator $\widetilde{B}$ is the
following linear combination of the previous $U_{V}(1)$ generator
$B=\frac{1}{3}{\rm diag}(1,1,1)$ and the broken diagonal generator
of the $SU_{c}(3)$ gauge group $T^{8}=\frac{1}{2\sqrt{3}}\,{\rm
diag}(1,1,-2)$:
\begin{equation}
\widetilde{B}=B-\frac{2\sqrt{3}}{3}T^{8}\ .  \label{residue}
\end{equation}
The quarks with color $1$ and $2$ are neutral under
$\widetilde{B}$ and consequently the condensate too
($\widetilde{B}$ is $\sqrt{2}\widetilde{S}$ of Ref.~\cite{CDS}).
The superconductive phase for $N_{f}=2$ possesses the same global
symmetry group of the confined Wigner-Weyl phase \cite{S}. In
Reference \cite{S}, it was shown that the low-energy spectrum, at
finite density, displays the correct quantum numbers to saturate
the 't~Hooft global anomalies \cite{tHooft}. It was also observed
that QCD at finite density can be envisioned, from a global
symmetry and anomaly point of view, as a chiral gauge theory
\cite{ball,ADS}. In Reference \cite{HSaS} it was then seen, by
using a variety of field theoretical tools, that global anomaly
matching conditions hold for any cold but dense gauge theory.

Massless excitations are protected by the aforementioned
constraints and dominate physical processes. The low-energy
theorems governing their interactions can be usefully encoded in
effective Lagrangians (like for cold and dilute QCD \cite{ChPT}).
It is possible to order the effective Lagrangian terms describing
the Golstone boson self interactions in number of derivatives. The
resulting theory for dilute QCD is named Chiral Perturbation
Theory \cite{ChPT}. Unfortunately this well defined scheme is not
sufficient for a complete description of hadron dynamics since new
massive hadronic resonances appear at relatively low energies.
{}For instance, the $\sigma$ or the vector $\rho$ and new
effective Lagrangian models of the type described in
\cite{effective Lagrangians} are needed.

 The dynamics of the Godstone bosons can be efficiently encoded in a
non-linear realization framework. Here, see \cite{CDS}, the
relevant coset space is $G/H$ with $G=SU_{c}(3)\times U_{V}(1)$
and $H=SU_{c}(2)\times \widetilde{U}_{V}(1)$ is parameterized by:
\begin{equation}
{\cal V}=\exp (i\xi ^{i}X^{i})\ ,
\end{equation}
where $\{X^{i}\}$ $i=1,\cdots ,5$ belong to the coset space $G/H$
and are taken to be $X^{i}=T^{i+3}$ for $i=1,\cdots ,4$ while
\begin{equation}
X^{5}=B+\frac{\sqrt{3}}{3}T^{8}={\rm
diag}(\frac{1}{2},\frac{1}{2},0)\ . \label{broken}
\end{equation}
$T^{a}$ are the standard generators of $SU(3)$. The coordinates
\begin{equation}
\xi ^{i}=\frac{\Pi ^{i}}{f}\quad i=1,2,3,4\ ,\qquad \xi
^{5}=\frac{\Pi ^{5}}{\widetilde{f}}\ ,
\end{equation}
via $\Pi $ describe the Goldstone bosons. The vevs $f$ and
$\widetilde{f}$ are expected, when considering asymptotically high
densities \cite{Son-S}, to be proportional to $\mu $.

${\cal V}$ transforms non linearly:
\begin{eqnarray}
{\cal V}(\xi )\rightarrow u_{V}\,g{\cal V}(\xi )h^{\dagger }(\xi
,g,u)h_{\widetilde{V}}^{\dagger }(\xi ,g,u)\ ,  \label{nl2}
\end{eqnarray}
with $u_{V}\in U_{V}(1)$, $g\in SU_{c}(3)$,  $h(\xi ,g,u)\in
SU_{c}(2)$ and $h_{\widetilde{V}}(\xi ,g,u)\in
\widetilde{U}_{V}(1)$.

It is convenient to define:
\begin{equation}
\omega _{\mu }=i{\cal V}^{\dagger }D_{\mu }{\cal V}\quad {\rm
with}\quad D_{\mu }{\cal V}=(\partial _{\mu }-ig_{s}G_{\mu }){\cal
V}\ ,
\end{equation}
with gluon fields $G_{\mu }=G_{\mu }^{m}T^{m}$ while $\omega $
transforms as:
\begin{eqnarray}
\omega _{\mu }\rightarrow && h(\xi ,g,u)\omega _{\mu }h^{\dagger
}(\xi ,g,u)+i\,h(\xi ,g,u)\partial {_{\mu }}h^{\dagger }(\xi
,g,u)\ \nonumber \\ &+&i\,h_{\widetilde{V}}(\xi ,g,u)\partial
_{\mu }h_{\widetilde{V}}^{\dagger }(\xi ,g,u).
\end{eqnarray}
Following \cite{CDS} we decompose $\omega _{\mu }$ into
\begin{equation}
\omega _{\mu }^{\parallel }=2S^{a}{\rm Tr}\left[ S^{a}\omega _{\mu
}\right] \quad {\rm and}\quad \omega _{\mu }^{\perp }=2X^{i}{\rm
Tr}\left[ X^{i}\omega _{\mu }\right] \ ,
\end{equation}
where $S^{a}$ are the unbroken generators of $H$ with
$S^{1,2,3}=T^{1,2,3}$, $S^{4}=\widetilde{B}\,/\sqrt{2}$. Summation
over repeated indices is assumed.

 To be able to include the
fermions in the picture we define:
\begin{equation}
\widetilde{\psi}={\cal V}^{\dagger }\psi \ ,  \label{mq}
\end{equation}
transforming as $\widetilde{\psi}\rightarrow
h_{\widetilde{V}}(\xi,g,u)h(\xi ,g,u)\widetilde{ \psi}$ and $\psi$
possesses an ordinary quark transformations (as Dirac spinor).

The simplest non-linearly realized effective Lagrangian describing
in medium fermions, the five gluons and their self interactions,
up to two derivatives and quadratic in the fermion fields is:
\begin{eqnarray}
{\cal L}=~ &&f^{2}a_{1}{\rm Tr}\left[ \,\omega _{0}^{\perp }\omega
_{0}^{\perp }-{\alpha }_{1}\vec{\omega}^{\perp
}\vec{\omega}^{\perp }\, \right] \nonumber \\ &+& f^{2}a_{2}\left[
{\rm Tr}\left[ \,\omega _{0}^{\perp }\,\right] {\rm Tr}\left[
\,\omega _{0}^{\perp }\,\right] -{\alpha }_{2}{\rm Tr}\left[
\,\vec{\omega}^{\perp }\,\right] {\rm Tr}\left[
\,\vec{\omega}^{\perp }\, \right] \right]  \nonumber \\
&+&b_{1}\overline{\widetilde{\psi }}i\left[ \gamma ^{0}(\partial
_{0}-i\omega _{0}^{\parallel })+\beta _{1}\vec{\gamma}\cdot \left(
\vec{ \nabla}-i\vec{\omega}^{\parallel }\right) \right]
\widetilde{\psi } \nonumber \\ &+& b_{2} \overline{\widetilde{\psi
}}\left[ \gamma ^{0}\omega _{0}^{\perp }+\beta _{2}
\vec{\gamma}\cdot \vec{\omega}^{\perp }\right] \widetilde{\psi }
\nonumber \\ &+&m_{M}\overline{\widetilde{\psi }^{C}}\gamma
^{5}(iT^{2})\widetilde{\psi }+ {\rm h.c.}\ , \label{cadusa}
\end{eqnarray}
where $\widetilde{\psi }^{C}=i\gamma ^{2}\widetilde{\psi }^{\ast
}$, $i,j=1,2 $ are flavor indices and
\begin{equation}
T^{2}=S^{2}=\frac{1}{2}\left(
\begin{array}{ll}
\sigma ^{2} & 0 \\ 0 & 0
\end{array}
\right) \ ,
\end{equation}
$a_{1},~a_{2},~b_{1}$ and $b_{2}$ are real coefficients while
$m_{M}$ is complex. The breaking of Lorentz invariance, following
\cite{CG} to the $O(3)$ subgroup has been taken into account by
providing different coefficients to the temporal and spatial
indices of the Lagrangian, and it is encoded in the coefficients
$\alpha $s and $\beta $s. For simplicity, the flavor indices are
omitted. {}From the last two terms, representing a Majorana mass
term for the quarks, we deduce that the massless degrees of
freedom are the $\psi _{a=3,i}$ which possess the correct quantum
numbers to match the 't~Hooft anomaly conditions \cite{S}. The
generalization to the electroweak processes relevant for the
cooling history of compact stars has been investigated in
\cite{CDS2001}.

The Lagrangian in Eq.~(\ref{cadusa}) together with the one
describing the relevant $SU_c(2)$ degree of freedom which we are
about to construct will be used elsewhere to derive the 2SC
equation of state.

\section{$SU_c(2)$ Glueball Effective Lagrangian}
\label{Glueball}

The $SU_c(2)$ gauge symmetry does not break spontaneously and it
is expected to confine. If the new confining scale is lighter than
the superconductive quark-quark gap the associated confined
degrees of freedom (light glueballs) can play, together with the
true massless quarks, as shown in \cite{OS} a relevant role for
the physics of Quark Stars featuring a 2SC superconductive surface
layer.

One would expect that below the scale $\Delta$, the heavy degrees
of freedom decouple and the low-energy theory is simply an
$SU_c(2)$ Yang Mills theory (together with the ungapped quarks);
with the new running coupling constant matched with the original
$SU_c(3)$ at the scale $\Delta$. However this is a misleading
argument. Indeed QCD at high chemical potential develops multiple
scales making it difficult to define a simple matching procedure.

Since the ungapped fermions are neutral with respect to the
$SU_c(2)$ - together with the diquarks built out of the quarks
carrying non trivial charge under $SU_c(2)$ - it would seem
natural for the medium to be transparent with respect to the
associated gluons. However, according to the findings in
\cite{rischke2k}, the medium does still lead to partial $SU_c(2)$
screening. In other words the medium is polarizable, i.e.,
acquires a dielectric constant $\epsilon$ different from unity (in
fact $\epsilon \gg 1$  in the 2SC case \cite{rischke2k}) leading
to an effectively reduced gauge coupling constant.

In general, a medium possesses a dielectric constant and a
magnetic permeability $\lambda\ne 1$. (Here, we note, that in the
approximations of \cite{rischke2k} $\lambda$ is still unity.) By
assuming locality the $SU_c(2)$ effective action takes the form
\cite{rischke2k}:
\begin{equation}
S_{eff}=\int \, d^4x
\left[\frac{\epsilon}{2}{\vec{E}^a}\cdot{\vec{E}^a}-\frac{1}{2\lambda}\vec{B}^a\cdot
\vec{B}^a\right] \label{sefu2}
\end{equation}
with $a=1,2,3$ and $E_{i}^a \equiv F^a_{0i}$ and  $B^a_{i}\equiv
\frac{1}{2}\epsilon_{ijk} F^a_{jk}$. Here one assumes an expansion
in powers of the fields and derivatives. The gluon speed in this
regime is $v=1/\sqrt{\epsilon \lambda}$.

In Reference \cite{rischke2k} the $\epsilon$ and $\lambda$ were
obtained studying the polarization tensor at asymptotically high
densities of the $SU_c(2)$ gluons and by finally expanding it in
powers of the momenta in order to get a local effective action.
Their results are:
\begin{equation}
\epsilon =1 + \frac{g_s^2 \mu^2}{18 \pi^2 \Delta^2}\ , \qquad
\lambda =1 \ . \label{el}
\end{equation}
Now, at asymptotically high densities, the gap $\Delta$ is
exponentially suppressed compared to the chemical potential $\mu$
\cite{Son}. Indeed $\Delta\propto \mu g_s^{-5} e^{-c/g_s}$ with
$c=3 \pi^2/\sqrt{2}$, while  $g_s$ is the $SU_c(3)$ coupling
constant evaluated at $\mu$. Equation (\ref{el}) than suggests
that a 2SC color superconductor can have a large positive
dielectric constant. This implies that the Coulomb potential
between $SU_c(2)$ color charges is reduced in the 2SC medium. The
fact that the medium easily polarizes can be intuitively
understood by recalling that Cooper pairs have a typical size of
the order of $1/\Delta$. There seems to be no effect on the
magnetic permeability and in the asymptotic high chemical
potential regime.

Clearly Eq.~(\ref{el}) is relevant to understand what happens to
the low-energy $SU_c(2)$ gluons. Unfortunately, although formally
correct and valuable, the perturbative results are very limited
when considering phenomenological applications since, according to
\cite{RS}, the results are quantitatively valid only for $\mu \gg
10^8$~MeV. Besides, the theory is believed to still confine and
hence $SU_c(2)$ glueballs like particles are expected to emerge.
These particles are light with respect to $\Delta$ and are shown
to play a relevant role in Quark Stars featuring a superconductive
2SC surface layer \cite{OS}. So, the low-energy $SU_c(2)$ theory
should be well represented by the effective Lagrangian describing
its hadronic low lying states: the light glueballs. This
Lagrangian has to be added to the one of Eq.~(\ref{cadusa})
\cite{CDS} and will be derived below.

A straightforward way to tackle the problem is to build the
$SU_c(2)$ energy stress tensor $\theta^{\mu \nu}$  whose trace is
related to a dilatation anomaly. We consider the theory at scales
lower than the gap.

The first step is to rescale the coordinates and the $SU_c(2)$
fields as follows:
\begin{eqnarray}
\hat{x}^0&=&\frac{x^0}{\sqrt{\lambda \epsilon}} \ , \qquad \hat{g}
= g_s\left(\frac{\lambda}{\epsilon}\right)^{\frac{1}{4}} \nonumber
\\ \hat{A}_0^a&=&
\lambda^{\frac{1}{4}}\epsilon^{\frac{3}{4}}A_0^a \ , \qquad
\hat{A}_i^a =
\lambda^{-\frac{1}{4}}\epsilon^{\frac{1}{4}}A_i^a\nonumber .\\
\label{drescaling}
\end{eqnarray}
In the limit $\lambda \rightarrow 1$ we recover the rescaling used
in \cite{rischke2k}. Here we consider a more general rescaling, by
not assuming $\lambda=1$, since for not too large chemical
potentials there is no guarantee that a small magnetic
permeability might not arise. If this were the case than note that
the coupling constant is sensitive to the ratio $\lambda/\epsilon$
which  is nevertheless much less than one (according to the
perturbative regime calculations).

Using the rescaled variables, the $SU_c(2)$ action becomes:
\begin{equation}
S_{SU(2)} = -\frac{1}{2} \int \, d^4 \hat{x}\, {\rm Tr} \left[
\hat{F}_{\mu \nu} \hat{F}^{\mu \nu} \right] \ , \label{effsu2}
\end{equation}
and $\hat{F}_{\mu,\nu}=\hat{\partial}_{\mu}\hat{A}_{\nu} -
\hat{\partial}_{\nu} \hat{A}_{\mu} +
i\,\hat{g}\left[\hat{A}_{\mu},\hat{A}_{\nu}\right]$  with
$\hat{A}_{\mu}=\hat{A}^a_{\mu}T^a$ and $a=1,2,3$.

The low-energy effective 3 gluon dynamics in the color
superconductor medium (with non-vanishing dielectric constant and
magnetic permeability) is similar to the in vacuum theory. The
expansion parameter is:
\begin{equation}
\hat{\alpha}=\frac{\hat{g}^2}{4\pi}=\frac{g_s^2}{4\pi}
\sqrt{\frac{\lambda}{\epsilon}} \ .
\end{equation}
Notice that $g_s$ is the $SU_c(3)$ coupling constant evaluated at
the scale $\mu$ while we now, following Ref.~\cite{rischke2k},
interpret $\hat{g}$ as the $SU_c(2)$ coupling at $\Delta$. The
matching of the scales is encoded in $\sqrt{\lambda/\epsilon}$.
Below $\Delta$ we use the action of Eq.~(\ref{effsu2}) to
investigate the $SU_c(2)$ properties.

Now we are ready to construct the glueball effective potential
valid to all orders in the loop expansion. This was achieved in
Ref.~\cite{SS} (in the vacuum case), by using the information of
the full, rather than just the one loop, beta function appearing
in the trace anomaly saturation procedure\cite{schechter}. The
explicit dependence on the full beta function of the theory
allowed \cite{SS} to investigate theories with large number of
flavors, relative to the number of colors, with nearby infrared
fix points.

The, in medium, anomaly-induced effective potential is based on
the trace anomaly arising from the rescaled theory written in
Eq.~(\ref{effsu2}):
\begin{equation}
\hat{\theta}_{\mu}^{\mu}=-\frac{\beta(\hat{g})}
{2\hat{g}}\hat{F}^{\mu\nu}_a\,\hat{F}_{\mu\nu;a}\equiv
\frac{2b}{v}\, H \ ,\label{trace}
\end{equation}
with $a=1,2,3$ and we have defined $\beta(\hat{g}) = -b {\hat
{g}^3}/16 \pi^2 $. At one loop $b=\frac{11}{3}N_c$ with $N_c=2$
the color number. $H$ is the composite field describing, upon
quantization, the scalar glueball \cite{schechter} in medium and
possesses mass-scale dimensions 4. The specific velocity
dependence is introduced to properly account for the velocity
factors.

The general nonderivative effective potential saturating the trace
anomaly is a solution of  \cite{schechter,joe,smrst}:
\begin{equation}
\hat{\theta}^{\mu}_{\mu}=4 H\frac{\delta \hat{V}}{\delta H} - 4
\hat{V} \ ,
\end{equation}
and is
\begin{equation}
\hat{V}=\frac{b}{2v} H\log \left[ \frac{H}{\hat{\Lambda}^4}
\right] \ , \label{potential}
\end{equation}
where $\hat{\Lambda}$ is some intrinsic scale associated with the
theory.

To the potential $\hat{V}$ one has still the freedom to add a non
derivative term proportional to $H$. Since this term does not
affect any of our conclusions it can be safely omitted. It is
worth mentioning that a similar type of potential was derived in
\cite{SSSusy} when breaking the ${\cal N}=1$ Super Yang-Mills
theory to ordinary Yang-Mills at the effective Lagrangian level.

In order to estimate $\hat{\Lambda}$ we consider the following one
loop relation:
\begin{eqnarray}
\hat{\Lambda}&=&\Delta \exp \left[-\frac{8\pi^2}{b_0
\hat{g}^2(\Delta)}\right] \nonumber \\&=&\Delta \exp
\left[-\frac{8\pi^2}{b_0 g_s^2(\mu)}{
\sqrt{\frac{\epsilon(\mu/\Delta)}{\lambda(\mu/\Delta)}}}\right]
\nonumber \\ &\simeq& \Delta \exp \left[-\frac{2\sqrt{2}\pi}{11}
\frac{\mu}{g_s(\mu)\Delta} \right]\ , \label{lambda}
\end{eqnarray}
with $b_0=22/3$ for $SU_c(2)$ and in the last step we considered
the  asymptotic solution of Ref.~\cite{rischke2k}, for convenience
reported in Eq.~(\ref{el}).

By using $\Lambda_{QCD}\simeq 300$ MeV, $\mu \simeq 500$ MeV and a
gap value of about $30$~MeV (roughly $2$ times  \cite{PR} the
superconductive critical temperature $\sim 15$ MeV estimated in
\cite{OS}) one gets $\hat{\Lambda} \simeq 1$~MeV. We recall that
this estimate relies on the one loop running and a reasonable
value of $\Lambda_{QCD}$ while we used the phenomenological value
for $\Delta$ \cite{OS}. This small value of $\hat{\Lambda}$
reinforces the need for the 2-color glueball effective Lagrangian
as a crucial ingredient of the low-energy effective action. Our
estimate is consistent with the one presented in \cite{rischke2k}
while further constraining $\hat{\Lambda}$ values.

The glueballs are light (with respect to the gap) and might barely
interact with the ungapped fermions. They are stable with respect
to the strong interactions unlike ordinary glueballs.

The potential in Eq.~(\ref{potential}) can be considered a zeroth
order model \cite{schechter,MS,SST} for a Yang-Mills theory, in
medium, in which the glueballs are the associated hadronic
particles. The potential has a minimum at
\begin{equation}\langle
H\rangle=\frac{\hat{\Lambda}^4}{e} \ ,
\end{equation}
at which point
\begin{equation}\langle\hat{V}\rangle=-\frac{b}{2v\,e}\hat{\Lambda}^4 \ .
\end{equation}
From Eq.~(\ref{trace}) this is seen to correspond to a
magnetic-type condensation of the glueball field $H$. The negative
sign of $\langle\hat{V}\rangle$ is consistent with the bag model
\cite{GJJK} in which a ``bubble" with $\langle\hat{V}\rangle=0$ is
stabilized against collapse by the zero point motion of the
particles within. For the zero density case a number of
phenomenological questions have been discussed using toy models
based on Eq.~(\ref{potential}) \cite{SST,GJJS,GJJS2}.

To be able to deduce further dynamical properties one needs a
kinetic term which does not affect the trace anomaly. A viable two
derivative trace invariant term (in the rescaled coordinates) is
\cite{GJJS}:
\begin{equation}
\hat{{\cal L}}_{kin}=v \frac{c}{2}\sqrt{b}\,
H^{-\frac{3}{2}}\hat{\partial}_{\mu} H \hat{\partial}^{\mu}H \ .
\end{equation}
Here $c$ is a positive dimensionless constant and the factor
$\sqrt{b}$ has been conveniently introduced. The previous
2-derivative term has scale dimensions four and hence does not
affect the trace anomaly equation. The complete simplest light
glueball action in the unrescaled coordinates for the, in medium,
Yang-Mill theory is:
\begin{eqnarray}
S_{G-ball}=\int
&d^4x&\left\{\frac{c}{2}\sqrt{b}\,H^{-\frac{3}{2}}\left[\partial^{0}
H
\partial^{0}H - v^2
\partial^iH
\partial^iH\right] \right.\nonumber \\ &&~~~ \left.  -\frac{b}{2}
H\log\left[\frac{H}{\hat{\Lambda}^4}\right] \right\} \ .
\label{G-ball}
\end{eqnarray}
Hence the glueballs move with the same velocity as the underlying
gluons in the 2SC color superconductor.

We define the mass-dimension one glueball field $h$ via:
\begin{equation}
H=\langle H\rangle e^{\frac{h}{F_h}} \ .
\end{equation}
By requiring a canonically normalized kinetic term for $h$ one
finds:
\begin{equation}
F_h^2=\frac{c}{\sqrt{2}} \sqrt{2b\langle H\rangle} \ .
\end{equation}
The glueball mass term is (obtained by expanding the Lagrangian up
to the second order in $h$):
\begin{equation}
M^2_h=\frac{\sqrt{b}}{2c}\sqrt{\langle
H\rangle}=\frac{\sqrt{b}}{2c\sqrt{e}} \hat{\Lambda}^2\ ,
\end{equation}
which is clearly of the order of $\hat{\Lambda}$ (estimated in the
MeV range from Eq.~(\ref{lambda})) since $c$ is a positive
constant of order unity. At large $N_c$ and for zero matter
density Yang-Mills theories one has $\langle
\hat{\theta}^{\mu}_{\mu}\rangle\sim {\cal O}\left(N^2_c\right)$
while $M^2_h \sim {\cal O}\left(N_c^0\right)$ which fixes $c\sim
{\cal O} \left(N_c\right)$ and hence $F^2_h\sim {\cal O}
\left({N^2_c}\right)$. Clearly, for large $N_c$ the glueball self
interactions are suppressed. However for theories at finite matter
densities  $N_c$ cannot be changed at will. Indeed a 2SC
superconductive phase is very sensitive to the actual number of
colors and flavors. Lattice results for $SU_c(2)$ Yang-Mills
theory \cite{MT} might be used to constraint the coefficients of
the effective Lagrangian which can then be used for extracting
dynamical results.

This completes the effective action for the $SU_c(2)$ glue-ball in
medium.

\section{$h\rightarrow \gamma \gamma$ process in the 2SC medium}
\label{decay} Once created, the light $SU_c(2)$ glueballs are
stable against strong interactions but not with respect to
electromagnetic processes. Indeed, in analogy to the case of the
$\pi^0$ decay into two photons at zero and high matter density for
the 3 flavor color superconductive case (CFL) \cite{CDS2001,HRZ},
the glueballs couple to two photons via virtual quark loops. More
specifically, the $\pi^0 \rightarrow \gamma\gamma$ process is a
direct consequence of the gauging of the global anomalies which
are seen to hold at finite matter density \cite{S,HSaS}, and the
explicit electromagnetic gauging is provided, for the CFL case, in
\cite{CDS2001}. On the other side the two-photon coupling of any
object which dominates the energy-momentum tensor at low energies
is based on the electromagnetic trace anomaly \cite{CE}. This has
been formulated at zero density by constructing a suitable
effective Lagrangian \cite{GJJS2} describing the ordinary glueball
decay into two photons. Mimicking the zero density case we modify
the trace-anomaly induced potential term as follows:
\begin{equation}
V= \frac{1}{4}\left[
\left(2bH+\widetilde{H}_{em}\right)\log\left[\frac{H}{\hat{\Lambda}^4}\right]\right]
\ ,
\end{equation}
where
\begin{equation}
\widetilde{H}_{em}=-\frac{\widetilde{e}^2}{24\pi^2}\left[\sum_{quarks
}\widetilde{Q}^2_{quarks}\right]
\widetilde{F}_{\mu\nu}\widetilde{F}^{\mu\nu} \ ,
\end{equation}
is the electromagnetic contribution to the trace anomaly, with
$\widetilde{F}_{\mu\nu}=\partial_{\mu}\widetilde{A}_{\nu}-
\partial_{\nu}\widetilde{A}_{\mu}$.
Here $\widetilde{A}_{\mu}$ is the in medium photon field
corresponding to the following massless linear combination of the
old photon and the eighth gluon \cite{charges,CDS2001}:
\begin{equation}
\widetilde{A}_{\mu }=\cos \theta _{Q}A_{\mu }-\sin \theta
_{Q}G_{\mu }^{8}\ ,
\end{equation}
with $\tan\theta _{Q}=e/(\sqrt{3}g_s)$. The new electric constant
is related to the in vacuum one via:
\begin{equation}
\widetilde{e}=e\,\cos \theta _{Q}\ .
\end{equation}
$\widetilde{Q}$ is the new electric charge operator associated
with the field $\widetilde{A}_{\mu }$:
\begin{equation}
\widetilde{Q}={\tau ^{3}\times {\mbox{\bf
1}}+\frac{\widetilde{B}-L}{2}}=Q\times {\mbox{\bf
1}}-\frac{1}{\sqrt{3}}{\mbox{\bf 1}\times }T^{8}\ ,
\label{newcharge}
\end{equation}
where $L=0$ is the lepton number, $\tau^3$ the standard Pauli's
matrix, $Q$ the quark matrix, the new baryon number
$\widetilde{B}$ is defined in Eq.~(\ref{residue}) and following
the notation of Ref.~\cite{CDS2001} we have ${\rm flavor}_{2\times
2}\times {\rm color}_{3\times 3}$.
 The quarks that acquires a mass term (i.e. the ones in the color
direction one and two) have half integer charges under
$\widetilde{Q}$  while the massless quarks (the ones in direction
three of color) have the ordinary proton and neutron charges in
units of $\widetilde{e}$. Hence:
\begin{eqnarray}
\sum_{quarks} \widetilde{Q}_{quarks}^2&=&{\rm
Tr}\left[\widetilde{Q}^2\right]=N_c {\rm Tr}{Q^2}+\frac{1}{3}=2 \
, \label{coeff}
\end{eqnarray}
with $N_c=3$ the underlying number of colors. Differently from the
$\pi^0 \rightarrow \gamma \gamma$ case, a part from a modified
electron coupling, one also finds an electromagnetic trace anomaly
coefficient which differs from the in vacuum case (corresponding
to the first term in Eq.~(\ref{coeff})). This, once again, shows
the special role played by the chiral anomalies.

The relevant Lagrangian term is:
\begin{equation} {\cal
L}_{h\gamma\gamma}=\frac{\widetilde{e}^2}{48\pi^2}
\frac{M_h}{\sqrt{2b\langle H  \rangle}} \left[\sum_{quarks}
\widetilde{Q}_{quarks}^2\right]
h\,\widetilde{F}_{\mu\nu}\widetilde{F}^{\mu \nu} \ ,
\end{equation}
leading to the following decay width of the glueballs into two
photons in medium:
\begin{eqnarray}
\Gamma\left[h\rightarrow \gamma\gamma\right]&=&\frac{\alpha^2}{576
\pi^3} \cos\theta_Q^4\left[\sum_{quarks}
\widetilde{Q}_{quarks}^2\right]^2 \, \frac{M_h^5}{2b\langle
H\rangle} \nonumber \\ &\approx& 1.2\times 10^{-2}
\cos\theta_{Q}^4 \left[\frac{M_h}{1~{\rm MeV}}\right]^5~{\rm eV} \
,
\end{eqnarray}
where $\alpha=e^2/4\pi \simeq 1/137$. {}For illustration purposes
we consider a glueball mass of the order of $1$~MeV which leads to
a decay time $\tau\sim~5.5\times~10^{-14}s$. We used
$\cos\theta_Q\sim~1$ since $\theta_Q \sim 2.5^{\circ}$ when
assuming for $\Lambda_{QCD}$ and the chemical potential the values
adopted in the previous section. While we are aware of the
possible contribution from other hadrons to the saturation of the
electromagnetic trace anomaly\cite{GJJS2}, here we assume it to be
dominated by the $SU_c(2)$ glueballs. In any case, it is hard to
imagine the photon decay process to be completely switched-off.

\section{Conclusion}
\label{conclusion} We  constructed the, in medium, effective
Lagrangian describing the light glueballs associated with the
unbroken and still confining $SU_c(2)$ color subgroup. This
Lagrangian has to be added to the one presented in \cite{CDS} and
constitutes a key ingredient for understanding the non
perturbative physics of 2 flavor color superconductivity. We have
shown that the light glueballs are unstable to photon decay and
estimated the, in medium, two photon decay rate.

This work shows that a consistent portion of the glue ($3/8$
 or $37.5\%$) filling the 2SC medium is very rapidly and efficiently
converted into electromagnetic radiation. The relevance of the
above in Quark Stars has been demonstrated in \cite{OS}. In
particular we showed that, if 2SC develops at the surface of such
stars (at the very early stage of their cooling history), glueball
formation and subsequent two-photon decay process provide the fuel
to power Gamma Ray Bursts. We discovered a plausible link between
Color Superconductivity and Gamma Ray Bursts with important
consequences to astrophysics and QCD \cite{OS}.

\acknowledgments

It is a pleasure for us to thank Joseph Schechter for interesting
discussions and encouragement.

\end{document}